\documentclass{article}

\PassOptionsToPackage{numbers, compress}{natbib}

\usepackage[final]{neurips_2019}

\usepackage[utf8]{inputenc} 
\usepackage[T1]{fontenc}    
\usepackage{hyperref}       
\usepackage{url}            
\usepackage{booktabs}       
\usepackage{amsfonts}       
\usepackage{nicefrac}       
\usepackage{microtype}      

\usepackage[pdftex]{graphicx}
\usepackage{subcaption}
\usepackage{amssymb}
\usepackage{url}

\title{Cognitive Assessment Estimation from Behavioral Responses in Emotional Faces Evaluation Task \\- AI  Regression Approach for Dementia Onset Prediction in Aging Societies  -}

\author{%
Tomasz M. Rutkowski$^\dag$, Masato S. Abe$^\dag$, Marcin Koculak$^\ddag$, and Mihoko Otake-Matsuura$^\dag$\\
 $^\dag$RIKEN Center for Advanced Intelligence Project (AIP), Tokyo, Japan\\
\url{http://aip.riken.jp/} \\
 \texttt{tomasz.rutkowski@riken.jp}\\
$^\ddag$Consciousness Lab, Institute of Psychology, Jagiellonian University, Krakow, Poland%
}

\begin{document}

\maketitle

\begin{abstract}
 We present a practical health-theme machine learning (ML) application concerning `AI for social good' domain for `Producing Good Outcomes' track. In particular, the solution is concerning the problem of a potential elderly adult dementia onset prediction in aging societies. 
The paper discusses our attempt and encouraging preliminary study results of behavioral responses analysis in a working memory-based emotional evaluation experiment. We focus on the development of digital biomarkers for dementia progress detection and monitoring. We present a behavioral data collection concept for a subsequent AI-based application together with a range of regression encouraging results of Montreal Cognitive Assessment (MoCA) scores in the leave-one-subject-out cross-validation setup. The regressor input variables include experimental subject's emotional valence and arousal recognition responses, as well as reaction times, together with self-reported education levels and ages, obtained from a group of twenty older adults taking part in the reported data collection project. The presented results showcase the potential social benefits of artificial intelligence application for elderly and establish a step forward to develop ML approaches, for the subsequent application of simple behavioral objective testing for dementia onset diagnostics replacing subjective MoCA.
\end{abstract}

\section{Introduction}
Dementia, especially the age-related memory decline, is one of the most significant global challenges in the $21^{st}$~century's mental well-being and social welfare. Worldwide, the increased longevity and mainly for elderly adults of above 65 years old, dementia numbers, and costs are rising~\cite{lancet2017dementia}. 
The Cabinet Office in Japan announces annual reports on an aging society to address the difficulty. United Nations Sustainable Development Goal ~$\#3$ entitled ``Good Health and Well-being'' also stresses a necessity to address the aging problem with a focus on healthy lives, and it promotes well-being for all at all ages.  
Recent approaches to dementia and Alzheimer's disease (AD) patient support suggest a necessity to develop personalized therapies relying not only on traditional pharmacological interventions but also on lifestyle modifications~\cite{theENDad2017} as well as cognitive support approaches~\cite{Reinhart:2019aa}. There is also a social expectation for the dementia early-onset prediction and subsequent prophylactics steps, as broadly discussed in~\cite{lancet2017dementia}. 
All the classical pharmacological and the novel `beyond-a-pill,' or the so-called `digital-pharma,' therapeutical interventions require trustful biomarkers, which would offer a comfortable alternative to more advanced in application brainwave-related techniques~\cite{dementiaBIOMARKER2018review,tomekNIPS2018}, which usually require a more clinical-level environment for a successful application. 
We propose a machine-learning (ML) approach, belonging to a broad spectrum of AI for the social or common good, which allows for automatic and objective estimation of a cognitive decline. A self-reported working-memory decline, the so-called subjective cognitive impairment (SCI), is one of the early biomarkers used in the medical community~\cite{Emrani2018}. A mild cognitive impairment (MCI), often preceding a dementia onset, is also characterized by emotional contagion~\cite{gutchess_2019} and hippocampus atrophy related spatial memory problems~\cite{lancet2017dementia}. 
On the other hand, there is no clear evidence about the working-implicit/procedural-memory impairment, and only the long-term equivalent is known to be unaffected in dementia subjects~\cite{slotnick_2017_chapter_implicit,gutchess_2019}. 
The contemporary methods for dementia diagnostics rely on pencil-and-paper subjective psychometric evaluations, for example, the Montreal Cognitive Assessment (MoCA)~\cite{moca2012}, the more complex physiological or imaging analyses~\cite{dementiaBIOMARKER2018review}, or massive multi-sensory datasets~\cite{chen2019developing}. The latter methods often require expensive devices, very long recording periods, or clinical settings.
We develop a machine-learning-based biomarker to replace the traditional MoCA, which utilizes behavioral responses in the spatial and working-implicit/procedural-memory testing task. Other work has been done predicting MCI using non-invasive data, but it has relied on subjective self-reports ~\cite{gutchess_2019}. We improve on their work by collecting behavioral data, which is less biased, when carrying out predictions.
We present an experimental and subsequent ML/AI behavioral response analysis approach in which we ask elderly subjects to learn a reasonable new emotional faces (Mind~Reading database~\cite{mindREADING}) evaluation skill using a two-dimensional map, a so-called emoji-grid~\cite{emojiGRID2018}, of arousal and valence scores, which is an effortless spatial- and implicit-working-memory task.
After a short training, the subjects perform a testing trial in which response time, arousal and valence user inputs together with self-reported age and education level features are used to train ML models as described in the following methods section. First, we conduct a pilot study with a small sample of university students, middle-age, high- $(\textrm{MoCA} > 25)$ and low-scoring ($20 \leqslant \textrm{MoCA} \leqslant 25$ in this study) on MoCA-scale elderly, as well as reference `super-normal' active-seniors of $80+$ years old. The pilot study produces encouraging results of reaction time (RT) and emotional valence/arousal (VA) behavioral responses, recorded with a touchpad, as biomarker candidates. The results of the pilot study inspire the main project in this paper with $20$ elderly MoCA-evaluated subjects. We report on promising regression results allowing for estimation of MoCA levels from behavioral responses and without the necessity for very subjects paper-and-pencil tests. 

\section{Methods}
We conducted experiments with human subjects with guidelines and approval of the 
RIKEN Ethical Committee for Experiments with Human Subjects in the Center for Advanced Intelligence Project (AIP). 
In the experimental session,
twenty elder participants (number of females~$=11$; mean age~$=76.5$ years old; age standard deviation~$=4.95$; recruited from Silver Human Resources Center) took part. All participants gave informed written consents, and they received a monetary gratification for their participation in the study.
Each subject experiment consisted of $72$~video presentation trials ($5\sim7$~seconds each) with $24$ different emotion categories~\cite{mindREADING}. Three different videos portrayed every emotion with actors differing in age, gender, and skin color. The order of the videos was randomized before the experiment but was the same for every participant. 
During the data recording experiments valence and arousal responses, as well as the reaction times were recorded together by the stimulus presentation application developed in a visual programming environment MAX by {\it Cycling~'74, USA}. We calculated absolute response errors and reaction times as, 
$v_{e}(i,s) = \left|v_{d}(i) - v_{t}(i,s)\right|, 
a_{e}(i,s) = \left|a_{d}(i) - a_{t}(i,s)\right|, 
r_t(i,s) = t_{t}(i,s) - t_o(i),$ 
with $s=1,\ldots,20$ identifying a participant in our study; $i=1,\ldots,72$ representing the video clip shown in a trial number $i$; $v_{e}(i,s)$ and $a_{e}(i,s)$ as valence and arousal errors related to emotional stimulus $i$ and subject $s$, respectively; $v_{d}(i)$ and $a_{d}(i)$ were the video clip assigned ground truth emotional scores from~\cite{mindREADING}; $v_{t}(i,s)$ and $a_{t}(i,s)$ the actual response inputs by a user number $s$ on a touchpad after the video clip number $i$, which reflected the learned emotion evaluation in the spatial- and working-memory task; $r_t(i)$ was a reaction time obtained as an interval between user response $t_{t}(i,s)$ and the $i^{th}$ video clip end at a timestamp $t_o(i)$. A single video clip evaluation feature vector $\textbf{F}_{i,s}$ related to video clip $i$ and participant $s$ for each evaluated next regressor in training and subsequent leave-one-subject-out cross-validation procedure has been built as, 
	$\textbf{F}_{i,s}=\left[v_{e}(i,s), a_{e}(i,s),r_t(i,s), e(s), g(s)\right],$ 
where $e(s) \in \{0,1\}$ denoted a self-reported education level, and $g(s)$ an age of $65\sim80$ years old in this study. 
A pairwise comparison scatter-plots of input features, together with linear regression fits, are summarized Figure~\ref{fig_pairplot}. 
We tested regressors available in {\it the scikit-learn} library version $0.21.3$~\cite{scikit-learn} for continues prediction of MoCA values characterizing cognition stages of the $20$ participants in our study using input features $\textbf{F}_{i,s}$ described previously. We implemented the leave-one-subject-out cross-validation procedure for our proposal evaluation.
The following methods and appropriate steps were implemented with default parameters set to as the scikit-learn, except for changes mentioned next to the methods: 
Huber regressor ({\bf HuberR}); 
linear regressor ({\bf linearR}); 
linear support vector regressor ({\bf linearSVR});
radial basis function support vector regressor ({\bf rbfSVR}): with a kernel coefficient $\gamma=1/7$ representing an inverse of feature vector $\textbf{F}_{i,s}$ length; 
polynomial support vector regressor ({\bf polySVR}): with a degree set to $3$, also here $\gamma=1/7$, an independent term in kernel function $coef_0=1.0$, $\epsilon=0.10$; 
random forest regressor ({\bf RFR}): with maximum depth set to $10$, and a number of estimators equal to $200$.
A binary classification trial for MCI $(\textrm{MoCA}\leqslant 25)$ versus normal cognition subjects resulted with median leave-one-subject out cross-validation accuracies of $99\%$ for LDA, LR, RFR; $98\%$ for linear SVM; $92\%$ for sigmoid SVM; $89\%$ and $83\%$ for RBF and polynomial SVMs, respectively (chance level of $50\%$). 

\section{Results and Conclusions}
\begin{figure}
	\centering
 	\begin{subfigure}[t]{0.60\textwidth}
		\includegraphics[width=\textwidth]{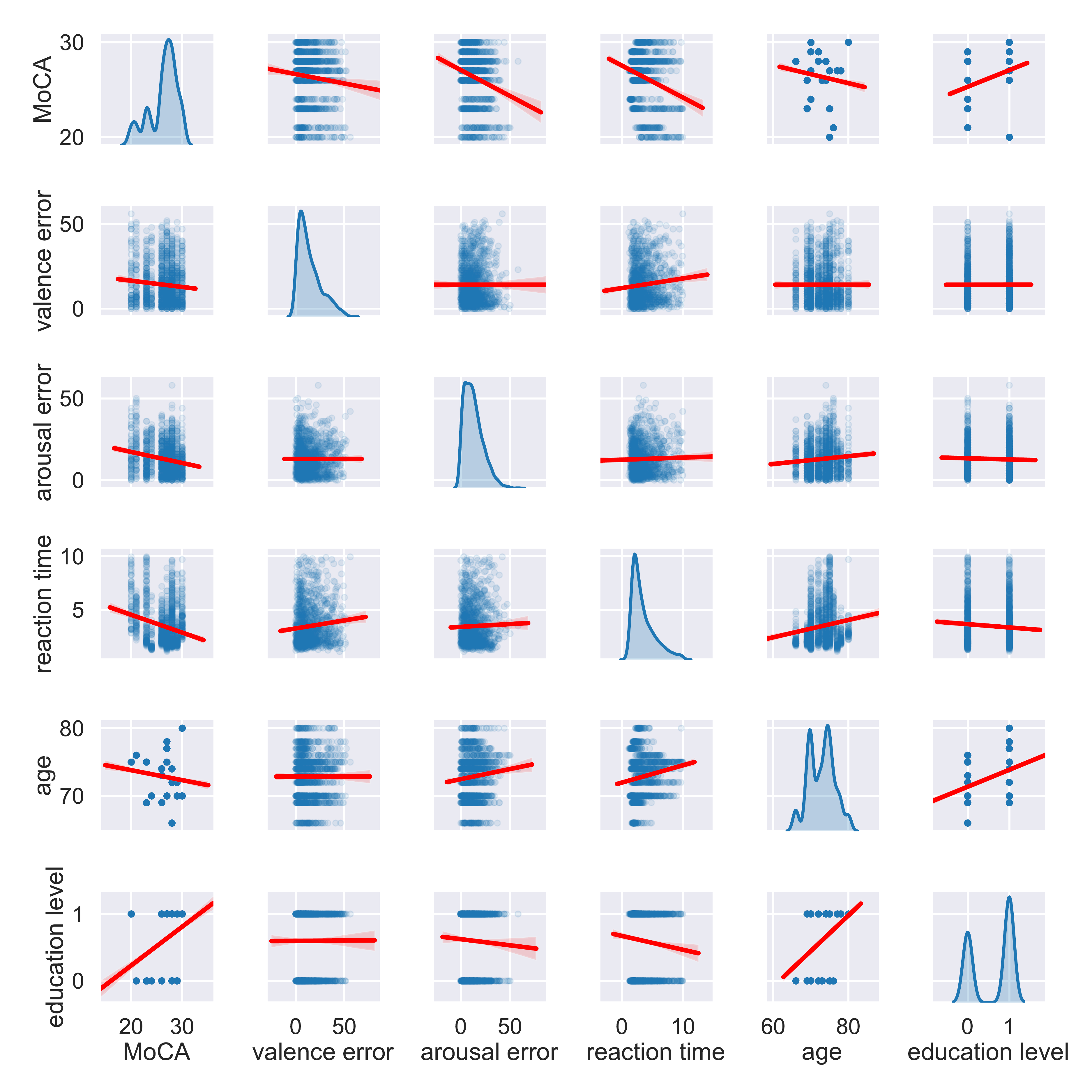} 
    		\caption{Pairwise plots of behavioral responses and MoCA}
    		\label{fig_pairplot}
  	\end{subfigure}
  	\begin{subfigure}[t]{0.385\textwidth}
		\includegraphics[width=\textwidth]{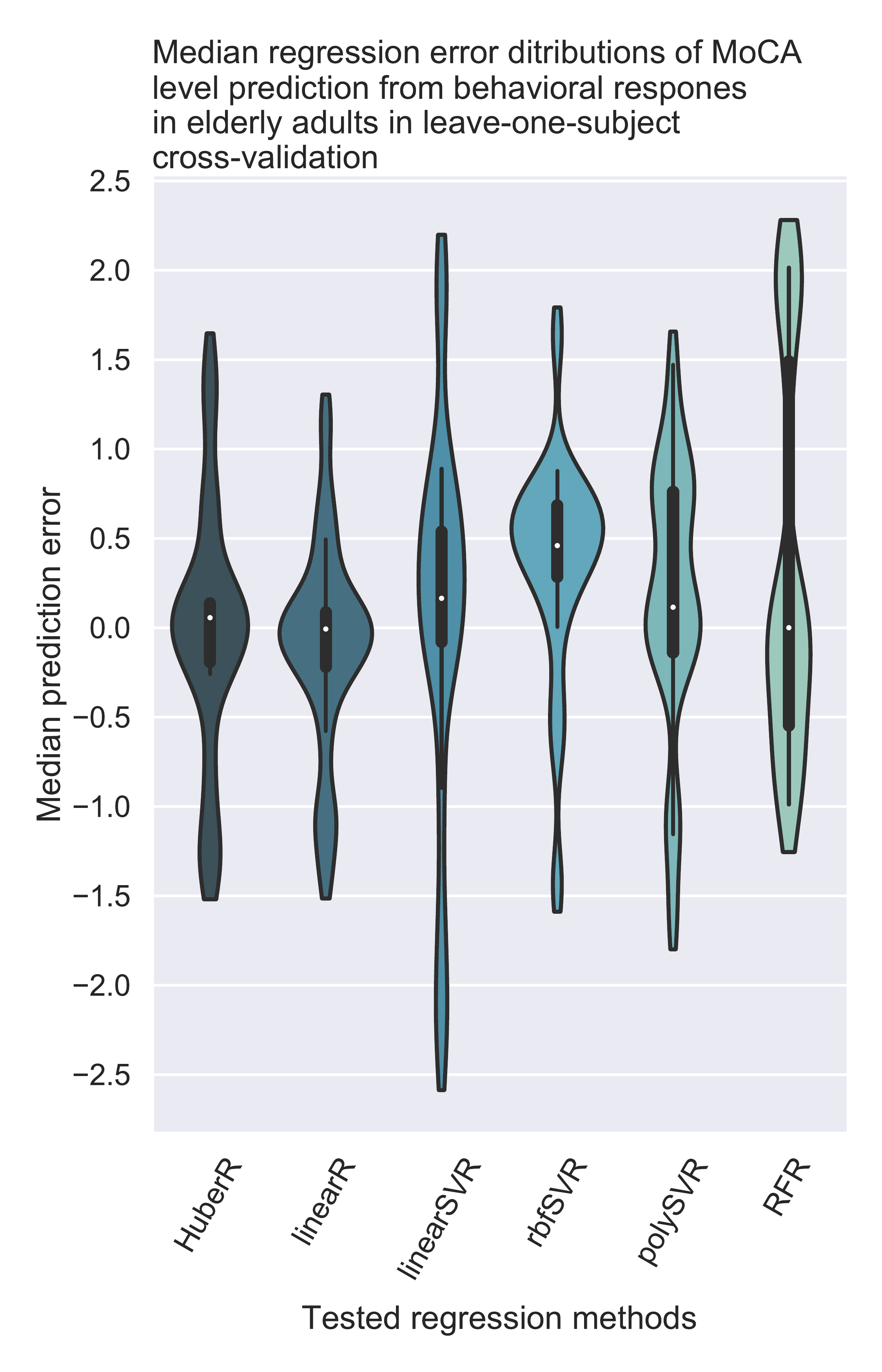} 
    		\caption{Regression median error results}
    		\label{fig_results}
  	\end{subfigure}
	\caption{Panel~(a) presents a collection of pair-plots showing relationships of output/target (MoCA) and input features used in the subsequent regression analysis. Red lines depict linear regression fits with shaded confidence intervals of the pairwise data distributions. Panel~(b) summarizes median regression errors from all the methods tested (naive mean regressor resulted in $+1.5$ median errors).}
	\label{fig_data}
\end{figure}

The current project resulted in encouraging results with a sample of $20$ older adults confirming a possibility of regression-based prediction of MoCA levels using only subject behavioral responses in the spatial- and implicit-working-memory task. 
The study results are summarized in the form of behavioral feature distributions and prediction errors in Figures~\ref{fig_pairplot}~and~\ref{fig_results}, respectively.
We expect that after collecting a more extensive database in near-future, a deep learning application would allow for an even more successful lowering of regression errors in MoCA predictions.
%
%
The study resulted in MoCA pathology prediction from behavioral responses in the spatial- and implicit-working-memory task of emotional valence and arousal levels estimation together with reaction times in simple video clips watching task. In the study involving older adults with known MoCA scores, we were able to evaluate several shallow learning regressors. The regression-based prediction of MoCA scores (usually $\textrm{MoCA}\leqslant 25$ has been considered as mild cognitive impairment (MCI) stage already, while above this threshold an elderly adult cognition has been evaluated as standard) in the simple emotional faces evaluation task resulted in robust and small errors as summarized in Figure~\ref{fig_results}.
The presented novel approach for behavioral responses in emotional faces evaluation task involving spatial- and implicit-working-memory-based new skill acquisition together with linear MoCA regression-based prediction results offer a step forward in research and development of novel dementia-related behavioral biomarkers for elderly adults, for whom possible early diagnosis of cognitive decline, as well as a life improvement, are essential. The successful application of such AI/ML-based dementia onset prediction shall lead to a healthcare cost lowering benefiting the aging societies.
We also acknowledge the potential limitations of the current approach as we only infer human-error-prone MoCA scores, which are only proxy estimators of dementia. AI-based dementia estimators, if used without proper evaluation, might also pose a danger of misuse or abuse; thus, proper ethical standards will need to be in place too.
In the next step of our research project, we plan to evaluate the developed methods with a larger sample of ordinary versus SCI/MCI, or even dementia diagnosed members of the society. We also plan to combine the proposed behavioral measures with neurophysiological, especially EEG and fNIRS signals for even more solid final classification. We also believe that future involvement of AI methods for fully interactive stimuli in closed-loop user behavior and brainwave monitoring shall lead to even more impactful results. 
\small

\begin{thebibliography}{13}
\providecommand{\natexlab}[1]{#1}
\providecommand{\url}[1]{\texttt{#1}}
\expandafter\ifx\csname urlstyle\endcsname\relax
  \providecommand{\doi}[1]{doi: #1}\else
  \providecommand{\doi}{doi: \begingroup \urlstyle{rm}\Url}\fi

\bibitem[Baron-Cohen(2004)]{mindREADING}
S.~Baron-Cohen.
\newblock \emph{Mind Reading - The Interactive Guide to Emotions}.
\newblock Jessica Kingsley Publishers, London, UK, 2004.
\newblock URL \url{http://www.jkp.com/mindreading/}.

\bibitem[Bredesen(2017)]{theENDad2017}
D.~Bredesen.
\newblock \emph{The End of Alzheimer's: The First Programme to Prevent and
  Reverse the Cognitive Decline of Dementia}.
\newblock Vermilion, 2017.

\bibitem[Chen et~al.(2019)Chen, Jankovic, Marinsek, Foschini, Kourtis,
  Signorini, Pugh, Shen, Yaari, Maljkovic, et~al.]{chen2019developing}
R.~Chen, F.~Jankovic, N.~Marinsek, L.~Foschini, L.~Kourtis, A.~Signorini,
  M.~Pugh, J.~Shen, R.~Yaari, V.~Maljkovic, et~al.
\newblock Developing measures of cognitive impairment in the real world from
  consumer-grade multimodal sensor streams.
\newblock In \emph{Proceedings of the 25th ACM SIGKDD International Conference
  on Knowledge Discovery \& Data Mining}, pages 2145--2155. ACM, 2019.

\bibitem[Emrani et~al.(2018)Emrani, Libon, Lamar, Price, Jefferson, Gifford,
  Hohman, Nation, Delano-Wood, Jak, Bangen, Bondi, Brickman, Manly, Swenson,
  and Au]{Emrani2018}
S.~Emrani, D.~J. Libon, M.~Lamar, C.~C. Price, A.~L. Jefferson, K.~A. Gifford,
  T.~J. Hohman, D.~A. Nation, L.~Delano-Wood, A.~Jak, K.~J. Bangen, M.~W.
  Bondi, A.~M. Brickman, J.~Manly, R.~Swenson, and R.~Au.
\newblock {Assessing working memory in mild cognitive impairment with serial
  order recall}.
\newblock \emph{J. Alzheimer's Dis.}, 61\penalty0 (3):\penalty0 917--928, 2018.
\newblock ISSN 18758908.
\newblock \doi{10.3233/JAD-170555}.

\bibitem[Gutchess(2019)]{gutchess_2019}
A.~Gutchess.
\newblock \emph{Cognitive and Social Neuroscience of Aging}.
\newblock Cambridge Fundamentals of Neuroscience in Psychology. Cambridge
  University Press, 2019.
\newblock \doi{10.1017/9781316026885}.

\bibitem[Horvath et~al.(2018)Horvath, Szucs, Csukly, Sakovics, Stefanics, and
  Kamondi]{dementiaBIOMARKER2018review}
A.~Horvath, A.~Szucs, G.~Csukly, A.~Sakovics, G.~Stefanics, and A.~Kamondi.
\newblock {EEG} and {ERP} biomarkers of {Alzheimer's} disease: a critical
  review.
\newblock \emph{Frontiers in Bioscience}, 23:\penalty0 183--220, 2018.

\bibitem[Julayanont et~al.(2012)Julayanont, Nasreddine, Brousseau, Borrie,
  Chertkow, and Phillips]{moca2012}
P.~Julayanont, Z.~Nasreddine, M.~Brousseau, M.~Borrie, H.~Chertkow, and
  N.~Phillips.
\newblock The {Montreal} cognitive assessment memory index score {(MOCA-MIS)}
  and total {MOCA} score to help predict mci conversion to {Alzheimer's}
  disease.
\newblock \emph{Alzheimer's \& Dementia: The Journal of the Alzheimer's
  Association}, 8\penalty0 (4):\penalty0 P372, 2012.

\bibitem[Livingston et~al.(2017)Livingston, Sommerlad, Orgeta, Costafreda,
  Huntley, Ames, Ballard, Banerjee, Burns, Cohen-Mansfield,
  et~al.]{lancet2017dementia}
G.~Livingston, A.~Sommerlad, V.~Orgeta, S.~G. Costafreda, J.~Huntley, D.~Ames,
  C.~Ballard, S.~Banerjee, A.~Burns, J.~Cohen-Mansfield, et~al.
\newblock Dementia prevention, intervention, and care.
\newblock \emph{The Lancet}, 390\penalty0 (10113):\penalty0 2673--2734, 2017.

\bibitem[Pedregosa et~al.(2011)Pedregosa, Varoquaux, Gramfort, Michel, Thirion,
  Grisel, Blondel, Prettenhofer, Weiss, Dubourg, Vanderplas, Passos,
  Cournapeau, Brucher, Perrot, and Duchesnay]{scikit-learn}
F.~Pedregosa, G.~Varoquaux, A.~Gramfort, V.~Michel, B.~Thirion, O.~Grisel,
  M.~Blondel, P.~Prettenhofer, R.~Weiss, V.~Dubourg, J.~Vanderplas, A.~Passos,
  D.~Cournapeau, M.~Brucher, M.~Perrot, and E.~Duchesnay.
\newblock Scikit-learn: Machine learning in {P}ython.
\newblock \emph{Journal of Machine Learning Research}, 12:\penalty0 2825--2830,
  2011.

\bibitem[Reinhart and Nguyen(2019)]{Reinhart:2019aa}
R.~M.~G. Reinhart and J.~A. Nguyen.
\newblock Working memory revived in older adults by synchronizing rhythmic
  brain circuits.
\newblock \emph{Nature Neuroscience}, 22\penalty0 (5):\penalty0 820--827, 2019.
\newblock \doi{10.1038/s41593-019-0371-x}.
\newblock URL \url{https://doi.org/10.1038/s41593-019-0371-x}.

\bibitem[Rutkowski et~al.(2018)Rutkowski, Zhao, Abe, and Otake]{tomekNIPS2018}
T.~M. Rutkowski, Q.~Zhao, M.~S. Abe, and M.~Otake.
\newblock {AI} neurotechnology for aging societies - task-load and dementia
  {EEG} digital biomarker development using information geometry machine
  learning methods.
\newblock In \emph{AI for Social Good Workshop at the Neural Information
  Processing Systems (NeurIPS) 2018}, pages 1--4, Montreal, Canada, December 8,
  2018.
\newblock URL
  \url{https://aiforsocialgood.github.io/2018/pdfs/track1/68-tomek_nips_2018_workshop-FINAL.pdf}.

\bibitem[Slotnick(2017)]{slotnick_2017_chapter_implicit}
S.~D. Slotnick.
\newblock \emph{Cognitive Neuroscience of Memory}, chapter Implicit Memory,
  pages 129--149.
\newblock Cambridge Fundamentals of Neuroscience in Psychology. Cambridge
  University Press, 2017.
\newblock \doi{10.1017/9781316026687}.

\bibitem[Toet et~al.(2018)Toet, Kaneko, Ushiama, Hoving, de~Kruijf, Brouwer,
  Kallen, and van Erp]{emojiGRID2018}
A.~Toet, D.~Kaneko, S.~Ushiama, S.~Hoving, I.~de~Kruijf, A.-M. Brouwer,
  V.~Kallen, and J.~B.~F. van Erp.
\newblock {EmojiGrid: A 2D Pictorial Scale for the Assessment of Food Elicited
  Emotions}.
\newblock \emph{Front. Psychol.}, 9:\penalty0 2396, nov 2018.
\newblock ISSN 1664-1078.
\newblock \doi{10.3389/fpsyg.2018.02396}.
\newblock URL
  \url{https://www.frontiersin.org/article/10.3389/fpsyg.2018.02396/full}.

\end{thebibliography}

\end{document}